# An Automated Group Key Authentication System Using Secret Image Sharing Scheme

**Dipak Kumar Kole [1], Subhadip Basu [2],**

Computer Science and Engineering Department,
MCKV Institute of Engineering, Liluah, Howrah-711204.
E-mail:{ [1] dipak_kole@rediffmail.com, [2] subhadip8@yahoo.com }

*Abstract*

*In an open network environment, privacy of group communication and integrity of the communication data are the two major issues related to secured information exchange. The required level of security may be achieved by authenticating a group key in the communication channel, where contribution from each group member becomes a part of the overall group key. In the current work, we have developed an authentication system through Central Administrative Server (CAS) for automatic integration and validation of the group key. For secured group communication, the CAS generates a secret alphanumeric group key image. Using secret image sharing scheme, this group key image shares are distributed among all the participating group members in the open network. Some or all the secret shares may be merged to reconstruct the group key image at CAS. A k-nearest neighbor classifier with 48 features to represent the images, is used to validate the reconstructed image with the one stored in the CAS. 48 topological features are used to represent the reconstructed group key image. We have achieved 99.1% classification accuracy for 26 printed English uppercase characters and 10 numeric digits.*

## 1.0. INTRODUCTION

With the near universal use of Internet in our day to day life, there emerge more and more group-oriented distributed applications. In the open network environment, application like multi-party computing, collaboration workspaces, video conferencing etc. are gaining immense popularity. However, there is a need for secured services to provide *privacy* of group communication and *integrity* of communication data. To achieve the required level of security, it is important to establish a common secret key for encrypting or signing the group communication data. This key is called the *group key*. In the communication channel suspect able to eavesdropping, each group member contributes their corresponding secret information share which becomes a part of overall group key.

A secret image sharing scheme, is a method for sharing a secret image among a set of participants. The secret image is encoded into *n* pieces, called *shares* and each share is given to a distinct participant (member). Certain *qualified* subsets of participants can recover the secret image by merging together their information, whereas *forbidden* subsets of participants can reveal no information. The specification scheme to identify the qualified and forbidden subsets is called *access structure*.

Authentication is all about establishing trust in human-human or human-computer interactions. It can be done by using something we know (e.g. a password), something we have (e.g. the key of an iron safe), or something we are (e.g. our facial attributes, fingerprint). All or some of these may also be confined to provide stronger authentication to enhance trustworthiness of the system. For example, a secured banking transaction system may ask its customer to produce the ATM card (something you have), type a personal identification number on the machine (something you know) and ask you to put your left thumb impression on a finger print sensor (something you are).

Previous research in this area has mostly concentrated on manual authentication through human visual system. In [1] N. Paul *et. al.* proposed a visual cryptographic scheme for internet based remote voting system. The system is developed using a simple 2 out of 2 image sharing scheme. Where, the original information image is subdivided into two secret share images. The information is retrieved visually through manual overlapping of the two image shares. The technique is not robust, as the secret information may be disclosed to all the viewers. Similarly, in [2] L.W.Hawkes *et. al.* has developed a system for generating up to 6 shares for a secure financial document image. Minimum 4 shares are required to reconstruct the original image. This system concentrated on communication of secret schemes over internet and reconstruction at receiver's end. But the reconstructed image is insecure at the receiver's end. Also the user at the receiver's end gets the knowledge of the secret image information.

The issue of authentication is addressed in [3] by W. Hailong *et. al.* They had proposed a scheme for authenticated group key agreement among the members of a group. They have used the ideas of 1-D-fasd encryption scheme and threshold cryptography to split the cryptographic operation among multiple users. However they have not concentrated on automatic authentication of the group key in a distributed environment.

In this paper we have developed a novel technique for automatic authentication of the group key in an open network environment. A *Central Administrative Server* (CAS) generates a secret group key for a specified group of *n* members. The CAS generates *n* secret shares from the secret group key image and distributes it among the members of the group. None of the members of the group or the manual administrator of the CAS has any knowledge about the secret key information. We have used k out of n secret image sharing scheme, such that at least *k* secret shares are required at the CAS to reconstruct the original group key image. Fig. 1 shows an illustration of the image share distribution scheme by CAS. In this automatic authentication system the major issue is to retrieve the meaning of the reconstructed group key image at the CAS. We have used a feature based recognition scheme to represent the reconstructed group key image.

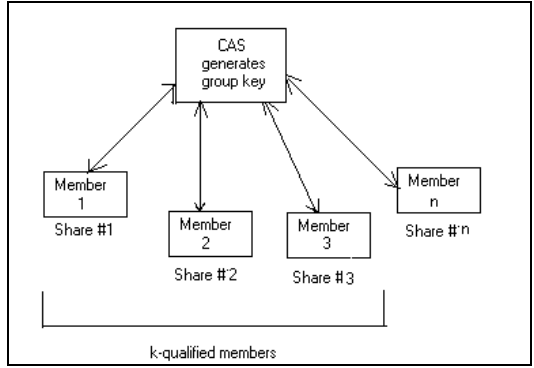

**Fig. 1. Secret image share distribution by CAS**

A *k*-nearest neighbor classifier is used to map the feature vector representation of the reconstructed image to the original group key information stored at CAS. In this automatic authentication process, if the reconstructed key image matches the desired group key information of the CAS, the *k* members of the group are allowed to access the secret.

On successful authentication, the members of a group can share a secret communication data among themselves, may view a secret image or information through CAS authenticate other users to join the group etc. The technique can be used in a high security environment sharing an open network, say internet. The major advantage of the technique is its simplicity yet robust security features.

The developed technique can be implemented for distribution of secret key information in an online banking system. Multiple users of a secured transaction account may be automatically authenticated at the CAS. Similarly, the technique is having immense potential for automatic user authentication in the secured military networks or in the research networks. The technique overcomes the problem of manual authentication in visual cryptographic schemes. Also the group key image remains secure at the receiver's end. Fig 2 shows a simple block diagram of the overall system architecture.

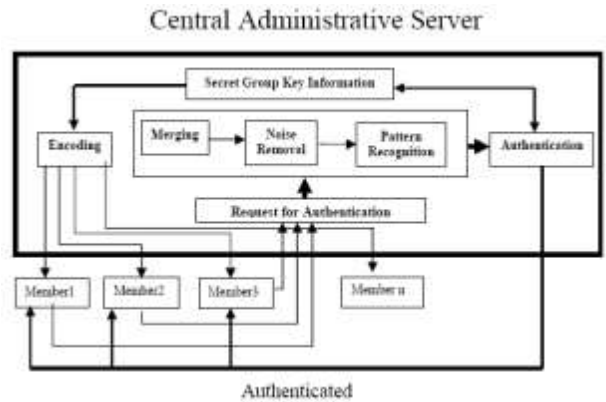

**Fig. 2. Schematic of the overall system architecture**

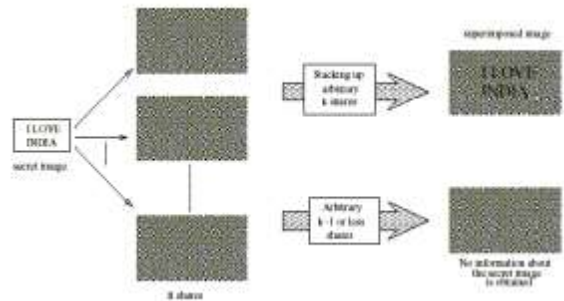

**Fig. 3. Illustration of the simple visual cryptographic scheme**

## 2.0. SECRET IMAGE SHARING SCHEME

In 1994, Naor and Shamir [4] proposed a new cryptographic scheme based on pixel level distribution of an image. They had introduced a new term of *visual cryptography* and established this as a method for encrypting such images of handwritten notes, pictures, graphical symbols as well as text stored as graphic images. The term *visual* came from the simple decoding technique with the help of human visual system. Fig. 3 shows the structure of a simple visual cryptographic scheme. However, in this paper we have deviated from the visual decoding technique to the automatic authentication process for the secret group key image.

## 2.1. The simple 2 out of 2 Secret Image Sharing Scheme

The basic 2 out of 2 visual cryptography model specifies how to encode a single pixel, and it would be applied for every pixel in the image to be shared. A pixel p is split into two pixels in each of the two shares, where each such pixel in the share is called sub pixel.

If p is white, then a coin toss is used to randomly choose one of the first two rows of the above figure. If p is black, then a coin toss is used to randomly choose one of the last two rows of the figure. Then the pixel is encrypted as two sub pixels in each of the two shares, as determined by the chosen row of the figure. Every pixel is encrypted using a new coin toss.

**Fig. 4. All possible distributions of black and white pixels in 2 out of 2 secret image sharing scheme**

If p is white, then a coin toss is used to randomly choose one of the first two rows of the above figure. If p is black, then a coin toss is used to randomly choose one of the last two rows of the figure. Then the pixel is encrypted as two sub pixels in each of the two shares, as determined by the chosen row of the figure. Every pixel is encrypted using a new coin toss.

Suppose we look at the two sub pixels, in the first share, corresponding to the pixel p in the secret image. One of these two sub pixels is black and other is white. Moreover, each of the two possibilities "black-white" and "white-black" is equally likely to occur, independent of whether the corresponding pixel in the secret image is black or white. Thus just by looking at the first share it is not possible to predict whether the sub pixels in the share correspond to a black or white pixel in the secret image. The same argument is applicable for the second share also. Since all the pixels in the secret image were encrypted using independent random coin flips, there is no information to be gained by looking at any group of pixels on a share, either. This demonstrates the security of the scheme.

When the shares are combined the original black pixel is viewed as black, however the original white pixel takes on a grey scale. Fig. 4 shows a pictorial representation of the simple 2 out of 2 secret image sharing scheme for a single pixel.

## 2.2. The Present Work

In the current work, a cryptography scheme is used using the idea of latin squares, as developed by A. Adhikari in [5]. Since a latin square exits for all orders and a special kind of latin square exits where the diagonals are in the natural order. Here we construct a (2, n) - VTS where n is of the form 3t, t>= 3. The important parameter of any VTS is its contrast i.e. the clarity with which the message becomes visible, and the pixel expansion i.e. number of sub pixels needed to encode one pixel of the original picture. We would like the contrast to be as large as possible and the pixel expansion as small as possible.

We know that, for any n of the form n=3t with t>=3, there exists a (2,n)- VTS with pixel expansion $n(n-3)/9 = t(t-1)$.

We first construct the basis matrices $S^0$ and $S^1$. To construct $S^1$, we need to construct the following matrix. First, consider a latin square $L = (l_{ij})_{t*t}$ of order t with entries from the set {1, 2,…t} with $l_{ij} = I$, V i= 1,2,…t. Then we consider the following arrangement

$$\begin{array}{ccc} \leftarrow t \rightarrow & \leftarrow t \rightarrow & \leftarrow t \rightarrow \\ 1\ 1\ …1 & 2\ 2\ …2 & …\ t\ t..t \\ 1\ 2\ …t & 1\ 2\ …t & …\ 1\ 2\ ..t \end{array}$$

After that, we write the rows of the latin square L consecutively as follows:

$$\begin{array}{cccc} 1\ 1\ …1 & 2\ 2\ …2… & t\ t…t \\ 1\ 2\ …t & 1\ 2\ …t… & 1\ 2…\ t \\ 1\ l_{12}…l_{1t} & l_{21}\ 2\ …\ l_{2t} & …\ l_{t1}\ l_{t2}…t \end{array}$$

Then from the above $3 * t^2$ matrix we delete all the columns with same entries. We delete exactly t columns. So the resulting matrix becomes a $3 * [t(t-1)]$ matrix. Let the resulting matrix be $N = (n_{ij})_{3*t(t-1)}$

$$\begin{array}{cccc} 1\ …1 & 2\ …2… & t\ t…t \\ 2\ …t & 1\ …t… & 1\ 2…\ t\text{-}1 \\ l_{12}…l_{1t} & l_{21}\ …\ l_{2t} & …\ l_{t1}\ l_{t2}…l_{tt\text{-}1} \end{array}$$

Since an element occurs exactly once in each row of L, this arrangement ensures each element i occurs exactly t-1 times in every row of N, i=1,2,…t. Now we write the elements of the above matrix in a different way to construct $M = (m_{ij})_{3*t(t-1)}$ as follows:

$$m_{1j} = (1, n_{1j}), j=1,…\ t(t-1).$$
$$m_{2j} = (1, n_{2j}), j=1,…\ t(t-1).$$
$$m_{3j} = (1, n_{3j}), j=1,…\ t(t-1).$$

Since the first coordinate has 3 choices and second coordinate has t choices, there will be total 3t symbols of the form (a, b) where a=1, 2, 3, and b=1, 2, t. Since each element $n_{ij}$ occurs exactly t-1 times in every row of N, each symbol (a, b) also occurs t-1 times in M. From the construction of M it is clear that symbol (a, b) for a given a, occurs in only one row of M. Now we rename the entries of the form (a, b) as $v_1, v_2,… v_{3t}$ and this remaining is done by the rule

$$(a, b) = v_{t(a-1)+b}, V\ a=1, 2, 3, b= 1, 2, …\ t.$$

Consider each column of M as one block $B_i$, i=1,…t(t-1). Each such block contains 3 entries and considers each entry as treatment. Thus, there are 3t treatments and t(t-1) blocks in all. The above construction ensures that any two treatments either occur together in exactly one block or they do not occur at all in any block. Now we construct the incident matrix

$$s^1 = 1 \text{ if } v_i \in B_j, \text{¥ } i= 1,2,… ,3t\ \ j=1,2,… ,t(t-1)$$
$$= 0, \text{ otherwise.}$$

$s^1$ is a 3t(t-1) matrix. Since each symbol occurs t-1 times in M, each row of $s^1$ contains exactly (t-1) 1's. We consider $s^1$ as the basis matrix for black pixel. The $3t*(t-1)$ basis matrix $s^0$ is realized by considering 3t identical n rows where first t-1 entries are 1 and remaining entries are 0.

## 3.0. AUTHENTICATION OF THE GROUP KEY IMAGE AT THE CAS

In any existing visual cryptographic scheme, the decoding is a trivial process. All the secret share images, distributed to n group members, are printed on transparent sheets. These transparent sheets are manually stacked one after another to retrieve the original secret image. The person involved retrieves the meaning of the reconstructed image. One of the major disadvantages of this scheme is the noise involved in the merged image. This is called the graying effect. Therefore, lossless reconstruction in secret image sharing scheme is a challenging problem.

In our work, we reconstruct the group key image at the CAS by simple ORing of pixel values of different image shares. The reconstructed image contains salt and peeper noise. We have designed an adaptive filter to remove the noise from the reconstructed image.

Isolated alphanumeric patterns are then segmented into isolated shapes from the key image and normalized into a 32x32 binary pixel matrix. A feature based recognition scheme is used, as discussed S. Basu *et. al.* in [6], to classify the isolated patterns into their designated class.

### 3.1. Design of the Adaptive Noise Filter

The filter design assumes that the image consists of black pixels representing data and white pixels representing white background only. From the discussions given in previous section we know that the image consists of white grains, termed as salt noise, in black regions and black grains, termed as pepper noise, in white regions.

**Table 1. Basic Algorithm for the adaptive filtering technique**

| Step 1 | For each pixel in the image |
|---|---|
| Step 2 | Calculate the number of black pixels and the number of white pixels in the area enclosed by the window. |
| Step 3 | Calculate the black to white pixels ratio. |
| Step 4 | If the ratio is lower than the white cut off the centre pixel is a white pixel. |
| Step 5 | If the ratio is higher than the black cut off the pixel is black. |
| Step 6 | If the ratio falls between the two cut offs the window size is increased and the process repeats until the upper limit on the window size is reached. |
| Step 7 | If after reaching the upper limit of the window size, still it is indecisive whether the pixel is white or black, leave the pixel as it is and moves to the next pixel. |

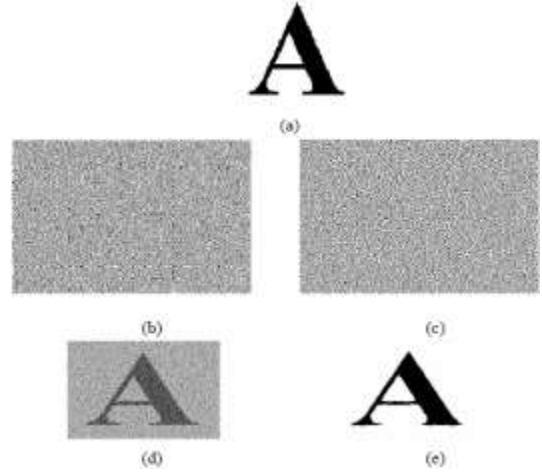

**Fig. 5. Illustration of the adaptive noise filter**
 (a) Original Image
 (b) Image of the secret share # 1
 (c) Image of the secret share # 2
 (d) **Reconstructed image corrupted with salt and peeper noise**
 (e) **Noise free image after applying the adaptive filter**

The basic algorithm for the filtering technique is shown in Table 1. Fig. 5 shows a sample alphabetic image distributed in two secret shares. On merging the shared images the reconstructed image contains salt and peeper noise. The adaptive filter removes the salt and peeper noise and retrieves the original image information.

### 3.2. Segmentation and Normalization

The adaptive noise filter removes the salt and peeper noise from the reconstructed group key image. Isolated shapes of each alphanumeric symbol need to be retrieval from the symbol string for subsequent feature extraction and classification. In the noise free key image we first compute the row wise black pixels count to identify the top and bottom bounding for the pattern images. A vertical column wise pixels scan is then performed to count the black pixels in each column. This identifies the left and the right boundaries for each pattern in the image.

The bounding box is thus identified by the top, bottom rows and left, right column boundaries to isolate each alphanumeric symbol from the key image.

### 3.3. Feature Extraction

A pattern is an arrangement of descriptors. The name feature is often used in the pattern recognition literature to denote a descriptor. A pattern class is a family of patterns that share the same common properties. Pattern classes are denoted by $\omega_1, \omega_2,\ldots, \omega_W$, where W is the number of classes. Pattern recognition by machine involves techniques for assigning patterns to their respective classes. Pattern vectors are represented by in the form, **x= ($x_1, x_2, \ldots, x_n$)**. Where, each component $x_i$ represents the ith descriptor and n is the total no of descriptors associated with that pattern.

In our scheme we have used 3 feature descriptors for pattern classification. These are
i) Black pixel/ total no of pixel ratio
ii) a) x -co ordinate of the centroid
    b) y -co ordinate of the centroid.

Each pattern image, to be recognized, is subdivided into 16 equal sub images. No of black pixel is counted in each sub image and it is divided by total no of pixel i.e. 64 and ratio is taken as a feature. For centroid calculation we have taken geometric moments of each part image. The centroids are given by

$$x_c = m_{10} / m_{00}$$
$$y_c = m_{01} / m_{00}$$

Where, $m_{pq}$ denotes the geometric moments

$$m_{pq} = \sum_x \sum_y x^p y^q f(x,y)$$

Here f(x,y) denotes the image function. The 3 feature values are computed for each of the 16 sub-images. Therefore 16x3 = 48 features are computed for each of the alphanumeric patterns. This 48 dimension feature vector is used for classification of 26 alphabet and 10 numeric classes.

### 3.4. Generation of Training Data

In our work, we have used 10 training samples for each pattern class. For the alphanumeric patterns we have used 26+10=36 classes and 36x10 = 360 training patterns. Training patterns are generated from standard fonts, *e.g.* Times New Roman, Courier, Arial etc. To generate the training patterns, images of each alphanumeric class, written in standard fonts, are first encoded in secret shared images and reconstructed by merging. Features are extracted from these training images after noise removal.

### 3.5. Design of the Pattern Classifier

After feature vector construction certain decision functions are taken into account. If there is a n-dimensional pattern vector and W pattern classes, as shown above, the basic problem in decision theoretic pattern recognition is to find W decision functions $d_1(x), d_2(x),\ldots, d_W(x)$ with the property that , if a pattern x belongs to class $\omega_i$ then, $d_i(x) > d_j(x)$ j=1,2,…W;

For the design of the classifier, we have used the *k*-nearest neighbor classification technique. The features are plotted in a N-dimensional feature space. In our scheme, we have used N=48 and *k*=1. The Euclidean distance between the calculated feature and saved feature in N-dimensional feature space are determined. Suppose there are two points $a=(a_1, a_2,\ldots, a_n)$ and $b=(b_1, b_2,\ldots b_n)$, then Euclidean distance between them is given by,

$$d_e(a,b) = \sqrt{\sum_{i=1}^{n} (b_i - a_i)^2}$$

Where n is the number of features. This is an extension of the Pythagorean Theorem to n-dimension. For each test pattern, we compute the distance metric with each of the training samples. If there are M training samples, then the minimum of the distances is obtained by

$$d_{min} = \min_{i=1}^{M} \{ d_i \}$$

Then, if $d_{min}$ corresponds to class $\omega_i$ then sample image is classified as $\omega_i$.

## 4.0. RESULTS AND DISCUSSION

We have developed a robust cryptographic scheme for high security in open network environments. In the current work, we have considered only black and white English alphanumeric symbols including 26 upper case and 10 numerals. Standard English fonts like Times New Roman, Arial, Courier etc are used for generation of the key image. Italics and designer fonts are not considered for the current experimentation. The 2 out of n secret sharing scheme is used to implement a simple, yet effective encryption scheme for the secret group key image. The technique can easily be enhanced for k out of n secret sharing scheme. It gives up two fold advantages. *Firstly,* any k-1 members cannot compute the desired functionality, so that the system becomes more secure. *Secondly,* an honest user who needs the cryptographic operation to be performed, need to contact at least k-1 other users.

In a distributed system, it avoids single point of failure and improves reliability of the system. Fig. 5 shows a sample secret group key image, generated secret share images and the reconstructed key image. We have considered 10 training samples for each of the 36 class alphanumeric patterns. For 500 test samples considered for experimentation, percentage recognition rate achieved is 99.10.

## 5.0. CONCLUSION

The present work can further be enhanced to incorporate general handwritten images of pictures containing grey level and colour information, maps etc. designer fonts and other language scripts may also be used in future with suitable training samples, better feature descriptors and effective classification techniques. The security can further be enhanced by incorporating an additional encoding mechanism to the original image, before the generation of multiple secret shares. Such that, the group members remain in complete dark about the group key information even if they merge all shares manually, i.e. without the knowledge of the CAS.